\documentclass[conference]{IEEEtran}
\IEEEoverridecommandlockouts
\usepackage{cite}
\usepackage{amsmath,amssymb,amsfonts}
\usepackage{algorithmic}
\usepackage{graphicx}     
\usepackage{textcomp}
\usepackage{xcolor}
\usepackage{url}
\usepackage{booktabs, multirow}
\DeclareMathAlphabet\mathcal{OMS}{cmsy}{m}{n}
\DeclareMathAlphabet\mathbfcal{OMS}{cmsy}{b}{n}

\def\BibTeX{{\rm B\kern-.05em{\sc i\kern-.025em b}\kern-.08em
    T\kern-.1667em\lower.7ex\hbox{E}\kern-.125emX}}

\usepackage{fancyhdr}
\fancypagestyle{firstpage}{
  \fancyhf{}
  \fancyhead[C]{\small To be presented at the \textit{IEEE International Workshop on Multimedia Signal Processing (MMSP)}, Shanghai, China, Sept. 26 - 28, 2022.}
}

\begin{document}

\title{Scalable Video Coding for Humans and Machines\\
\thanks{This work was partially supported by the Natural Sciences and Engineering Council (NSERC) of Canada.}
}

\author{\IEEEauthorblockN{Hyomin Choi and Ivan V. Baji\'{c}}
\IEEEauthorblockA{\textit{School of Engineering Science, Simon Fraser University}\\
Burnaby, BC, Canada \\
}
}

\maketitle

\begin{abstract}
Video content is watched not only by humans, but increasingly also by machines. For example, machine learning models analyze surveillance video for security and traffic monitoring, search through YouTube videos for inappropriate content, and so on. In this paper, we propose a scalable video coding framework that supports machine vision (specifically, object detection) through its base layer bitstream and human vision via its enhancement layer bitstream. The proposed framework includes components from both conventional and Deep Neural Network (DNN)-based video coding. The results show that on object detection, the proposed framework achieves 13-19\% bit savings compared to state-of-the-art video codecs, while remaining competitive in terms of MS-SSIM on the human vision task.
\end{abstract}


\thispagestyle{firstpage}

\section{Introduction}
\label{sec:introduction}
Video analytics is an essential technology for various applications such as traffic monitoring, visual surveillance, and autonomous navigation.
Automated machine vision pipelines increasingly analyze video streams  uploaded to the cloud. If one is only interested in (machine-based) visual analysis, pre-computed features can be compressed and transmitted~\cite{duan2020video}, instead of the full video. There are standard codecs to compress computed features, either handcrafted or neural network-based~\cite{cdvs_std, cdva_std}, although these were developed prior to the current wave of interest on the topic. A problem with these approaches, however, is that when human viewing is needed, input video must also be coded and transmitted, and the overall system becomes less efficient.  

A recent trend in image/video coding is to utilize deep neural networks (DNNs) to replace either specific units within conventional codecs, or even the whole codec. Over the past few years, DNNs have made inroads in this area, often demonstrating promising coding results compared to fully-engineered conventional approaches~\cite{balle2017iclr,balle2018variational,cheng2020image,Dandan2021_advances, dnn_video_coding_tcsvt}. However, most DNN-based codecs have focused on compression for human vision, just like traditional codecs. At the same time, there are many DNN-based vision analysis methods~\cite{zhang2019deep} but they are usually developed without regard for compression. To establish a consolidated framework that supports both human and machine vision, a standardization activity, MPEG-VCM (Video Coding for Machines)~\cite{vcm_call_for_evidence}, has recently been initiated. Several recent proposals for image coding~\cite{hu2020towards, liu2021semantics,  choi2022scalable} examined scalable compression for multiple tasks. In these methods, the base layer features are used to perform machine vision. With additional information in the enhancement layer, these methods also support high-quality input reconstruction for human vision. Meanwhile,~\cite{Choi2018NearLosslessDF, cuboidal_icip21} tried to recover the input image directly from features, without an additional bitstream:~\cite{Choi2018NearLosslessDF} from intermediate-layer activations of YOLOv2~\cite{YOLO2}, and~\cite{cuboidal_icip21} from cuboidal features targeted at YOLOv2. If $\mathbf{X}$ denotes the input image, $\mathbfcal{Y}$ the latent space features, $\widehat{\mathbf{X}}$ the reconstructed image, and $T$ the machine task output, then a typical machine vision pipeline can be described by a Markov chain $\mathbf{X} \to \mathbfcal{Y} \to \widehat{\mathbf{X}} \to T$. Applying the data processing inequality~\cite{Cover_Thomas_2006} to this chain, we obtain $I(\mathbfcal{Y};\widehat{\mathbf{X}}) \geq I(\mathbfcal{Y};T)$, where $I(\cdot \,;\cdot)$ is the mutual information, suggesting that less information (fewer bits) is needed for machine vision than for input reconstruction.\footnote{Intuitively, we don't need the details of every pixel in order to detect objects.} This agrees with the scalable approaches, where both base and enhancement layers are used for input reconstruction, but only the base layer (i.e., fewer bits) for machine vision.     
  
In this paper, we develop the first (to our knowledge) video codec for human and machine vision based on the concept of \textit{latent-space scalability}~\cite{choi_lss_icip2}. The codec utilizes multi-task DNN-based compression for intra coding and a combination of DNN and conventional techniques for inter-frame coding.    
In Section~\ref{sec:proposed}, we present the proposed video coding system, along with the explanation of its building blocks for intra and inter coding.  
Experimental results are presented in Section~\ref{sec:ch6_experimental_results}, followed by conclusions in Section~\ref{sec:ch6_conclusion}.

\section{Proposed methods}
\label{sec:proposed}
  
\begin{figure}[t]
    \centering
    \begin{minipage}[b]{1\linewidth}
    \centering
    \includegraphics[width=\textwidth]{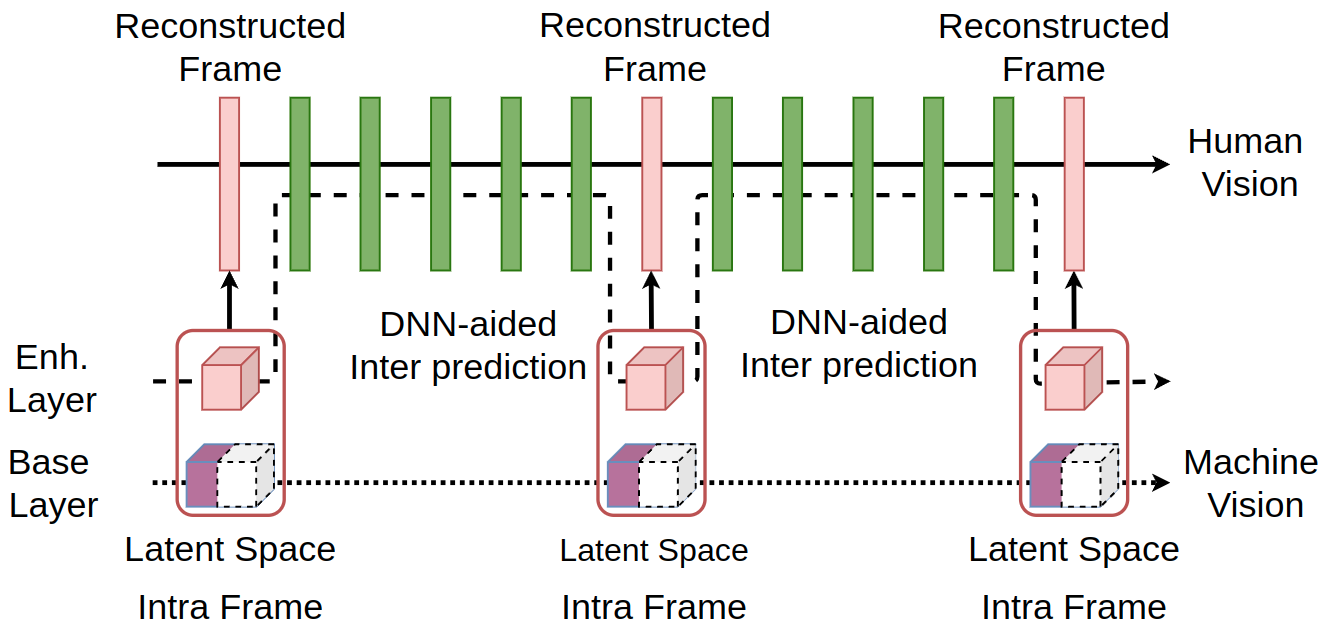}
    \end{minipage}
\caption{Proposed video coding for humans and machines.}
\label{fig:GoP_structure}
\vspace{-5pt}
\end{figure}

\begin{figure*}
    \centering
    \includegraphics[width=0.75\textwidth]{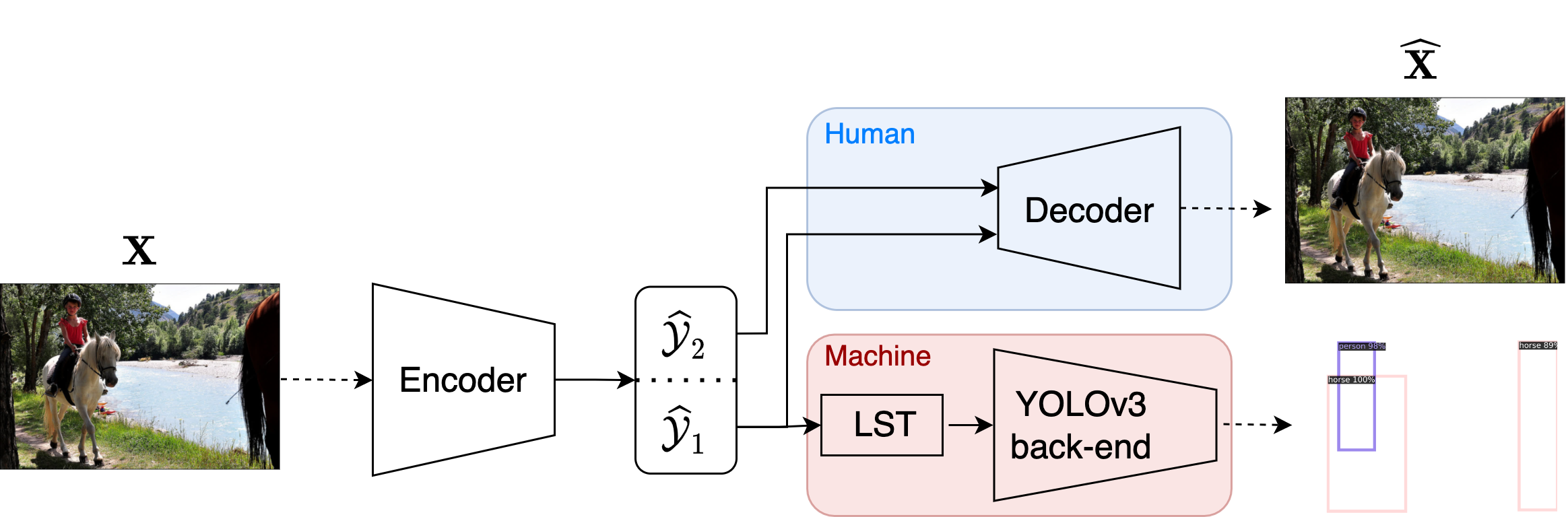}
    \caption{Scalable multi-task intra coding.}
    \label{fig:intra_coding}
    \vspace{-5pt}
\end{figure*}

\subsection{Group-of-Pictures structure}
\label{sec:GoP}
The Group-of-Pictures (GoP) structure of our proposed video coding system is shown in Fig.~\ref{fig:GoP_structure}. The GoP consists of an intra-coded frame (shown as pink in Fig.~\ref{fig:GoP_structure}) and a number of subsequent inter-coded  frames (shown as green in Fig.~\ref{fig:GoP_structure}). Intra frames are coded in a multi-task scalable manner using the concept of latent-space scalability, as detailed below in Section~\ref{sec:intra}. The base layer of an intra-frame bitstream supports machine vision, in our case object detection by YOLOv3~\cite{Redmon2018_yolov3}. When only machine vision is needed, only the base layer bitstream needs to be decoded. The enhancement layer provides additional information for high-quality input reconstruction for human viewing. Inter frames are exclusively used for human viewing. Hence, inter-frame bitstrams are considered a part of the enhancement layer. 

The intended application of such a system is as follows. Under normal operation, only the machine vision task is active. For example, a cloud-based automated analysis pipeline is monitoring an airport lounge through a surveillance camera. The camera sends the base-layer bitstream to the cloud, where the machine vision pipeline performs object detection. When a situation of interest is detected -- for example, unattended luggage\footnote{A piece of luggage without a human detected near it.} -- enhancement layer is requested by the cloud. This means that both the base and the enhancement layer bitstream of the next intra frame are sent to the cloud, together with subsequent inter frames. Security personnel can then view the scene and decide when to switch back to machine-only mode. 

This example was given as a specific illustration of the intended use of the proposed system. The use is not limited to public area surveillance -- similar situations arise in traffic monitoring, 
smart home security, and so on.  

\subsection{Multi-task scalable intra coding}
\label{sec:intra}
The structure of the multi-task scalable intra frame encoder follows that of
~\cite{choi2022scalable}, and is illustrated in Fig.~\ref{fig:intra_coding}. Intra frame $\mathbf{X}$ is encoded into a latent representation $\widehat{\mathbfcal{Y}}$ as:
\begin{equation}
    \mathbfcal{Y} = g_a(\mathbf{X}), \qquad \widehat{\mathbfcal{Y}} = \mathcal{Q}(\mathbfcal{Y}),
\end{equation}
where $g_a(\cdot)$ is a trainable analysis transform~\cite{balle2017iclr,balle2018variational,cheng2020image} and $\mathcal{Q}(\cdot)$ is the quantizer, implemented as rounding to the nearest integer. The architecture of the encoder can be taken from any end-to-end trainable image codec; we have used the architecture from~\cite{cheng2020image}. The main difference is that we retrain the encoder to produce a structured latent space $\widehat{\mathbfcal{Y}}=\{\widehat{\mathbfcal{Y}}_1,\widehat{\mathbfcal{Y}}_2\}$, such that $\widehat{\mathbfcal{Y}}_1$ contains the information needed for object detection (base layer), and $\widehat{\mathbfcal{Y}}_2$ contains enhancement information which, together with $\widehat{\mathbfcal{Y}}_1$, enables reconstruction of the input frame. Out of the 192 channels in $\widehat{\mathbfcal{Y}}$, 128 are assigned to $\widehat{\mathbfcal{Y}}_1$ and the remaining 64 to $\widehat{\mathbfcal{Y}}_2$.

From such latent representation, two decoders -- one for machine vision, the other for human viewing -- can efficiently reconstruct the required information. The object detection decoder decodes only $\widehat{\mathbfcal{Y}}_1$:
\begin{equation}
    \mathbfcal{\widetilde{F}}=\text{LST}(\widehat{\mathbfcal{Y}}_1), \qquad T = \mathcal{O}(\widetilde{\mathbfcal{F}}),
    \label{eq:machine_decoding}
\end{equation}
where LST stands for the Latent Space Transform~\cite{choi_lss_icip2} that maps the encoder's latent space into a latent space of the object detection network, and $\mathcal{O}(\cdot)$ is the back-end of the object detection network producing detection output $T$. In our case, we have chosen layer 13 of YOLOv3 as the target latent space, so  $\mathcal{O}(\cdot)$ consists of all YOLOv3 layers after layer 13. Several possible architectures for the LST were presented in~\cite{choi_lss_icip2,choi2022scalable}, here we have used the one from~\cite{choi2022scalable}.

Meanwhile, the decoder  for human viewing decodes the entire latent space $\widehat{\mathbfcal{Y}}=\{\widehat{\mathbfcal{Y}}_1,\widehat{\mathbfcal{Y}}_2\}$,
\begin{equation}
    \widehat{\mathbf{X}} = \mathcal{D}(\widehat{\mathbfcal{Y}}),
    \label{eq:human_decoder}
\end{equation}
where $\mathcal{D}$ is the decoder and and $\widehat{\mathbf{X}}$ is the approximation to the original input frame $\mathbf{X}$. The architecture of the decoder $\mathcal{D}$ can be taken from the same end-to-end codec where the encoder came from; in our case, this is~\cite{cheng2020image}.

Structuring of the latent space is achieved by training all components of the encoder in Fig.~\ref{fig:intra_coding}, except $\mathcal{O}$, the object detection back-end, jointly using the loss function
\begin{equation}
    \mathcal{L} = R + \lambda\cdot \text{MSE}(\mathbf{X},\widehat{\mathbf{X}}) + \lambda\cdot \gamma \cdot \text{MSE}(\mathbfcal{F},\widetilde{\mathbfcal{F}}), 
    \label{eq:loss_function}
\end{equation}
where $R$ is a rate estimate (obtained using the entropy model from~\cite{cheng2020image}), $\lambda$ controls the trade-off between rate and distortion, and $\gamma$ controls the trade-off between latent-space distortion for object detection, $\text{MSE}(\mathbfcal{F},\widetilde{\mathbfcal{F}})$, and pixel-domain distortion for input reconstruction, $\text{MSE}(\mathbf{X},\widehat{\mathbf{X}})$. Since $\widetilde{\mathbfcal{F}}$ is computed only from $\widehat{\mathbfcal{Y}}_1$, as shown in~(\ref{eq:machine_decoding}), by gradient-based training, information related to object detection will be steered only into $\widehat{\mathbfcal{Y}}_1$. Meanwhile, since $\widehat{\mathbf{X}}$ is obtained from the entire latent space $\widehat{\mathbfcal{Y}}$, as shown in~(\ref{eq:human_decoder}), information related to input reconstruction will be spread throughout the entire latent space, including $\widehat{\mathbfcal{Y}}_2$.

\subsection{Inter-frame coding}
\label{sec:inter}
\begin{figure}[t]
    \includegraphics[width=\columnwidth]{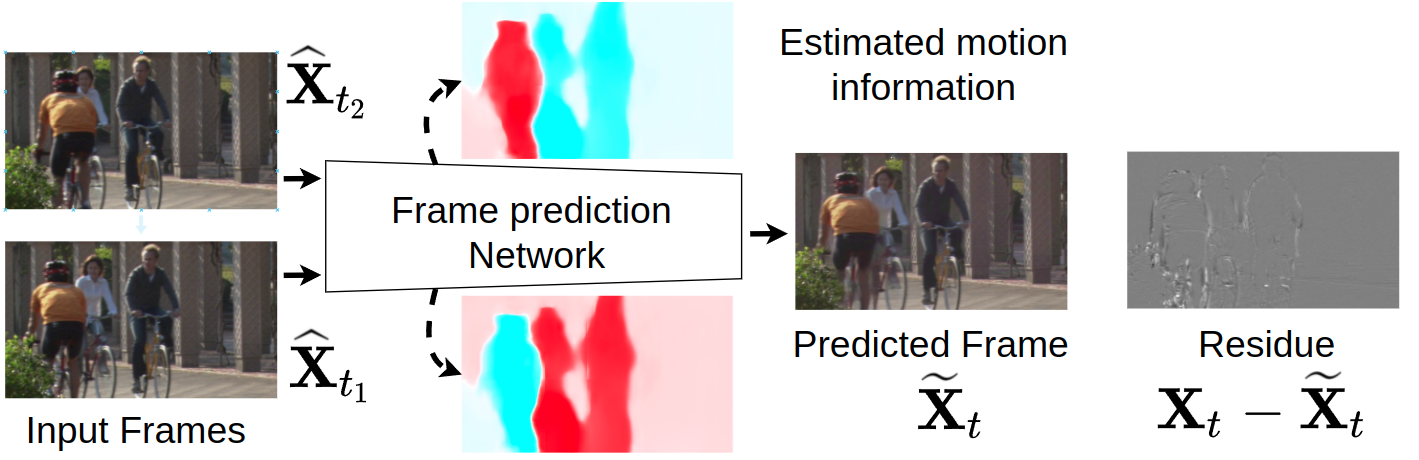}
\caption{Bi-directional affine transformation-based deep frame prediction from~\cite{choi2021affine}.}
\vspace{-.2cm}
\label{fig:frame_prediction}
\end{figure}

Inter-frame coding is based on High Efficiency Video Coding (HEVC)~\cite{hevc_std}, specifically HEVC test model HM-16.20.\footnote{\url{http://hevc.hhi.fraunhofer.de/svn/svn_HEVCSoftware/tags/HM-16.20+SCM-8.8}} Besides conventional HEVC coding tools, we also integrate DNN-based affine frame prediction from~\cite{choi2021affine} into the coding pipeline. The operation of this network is illustrated in Fig.~\ref{fig:frame_prediction}. The network takes two previously coded frames $\widehat{\mathbf{X}}_{t_1}$ and $\widehat{\mathbf{X}}_{t_2}$ as input, and predicts a half-way frame $\widetilde{\mathbf{X}}_t$ between the input frames, where $t_1<t<t_2$. Internally, the network estimates motion between the two input frames and the desired output frame, as visualized in the figure, along with adaptive filter kernels. With computed motion and filter kernels, the network produces an estimated frame such that the residual signal to code is minimized. The reader is referred to~\cite{choi2021affine} for details about the prediction network.

\begin{figure}[t]
    \centering
    \begin{minipage}[b]{0.95\linewidth}
    \centering
    \includegraphics[width=\textwidth]{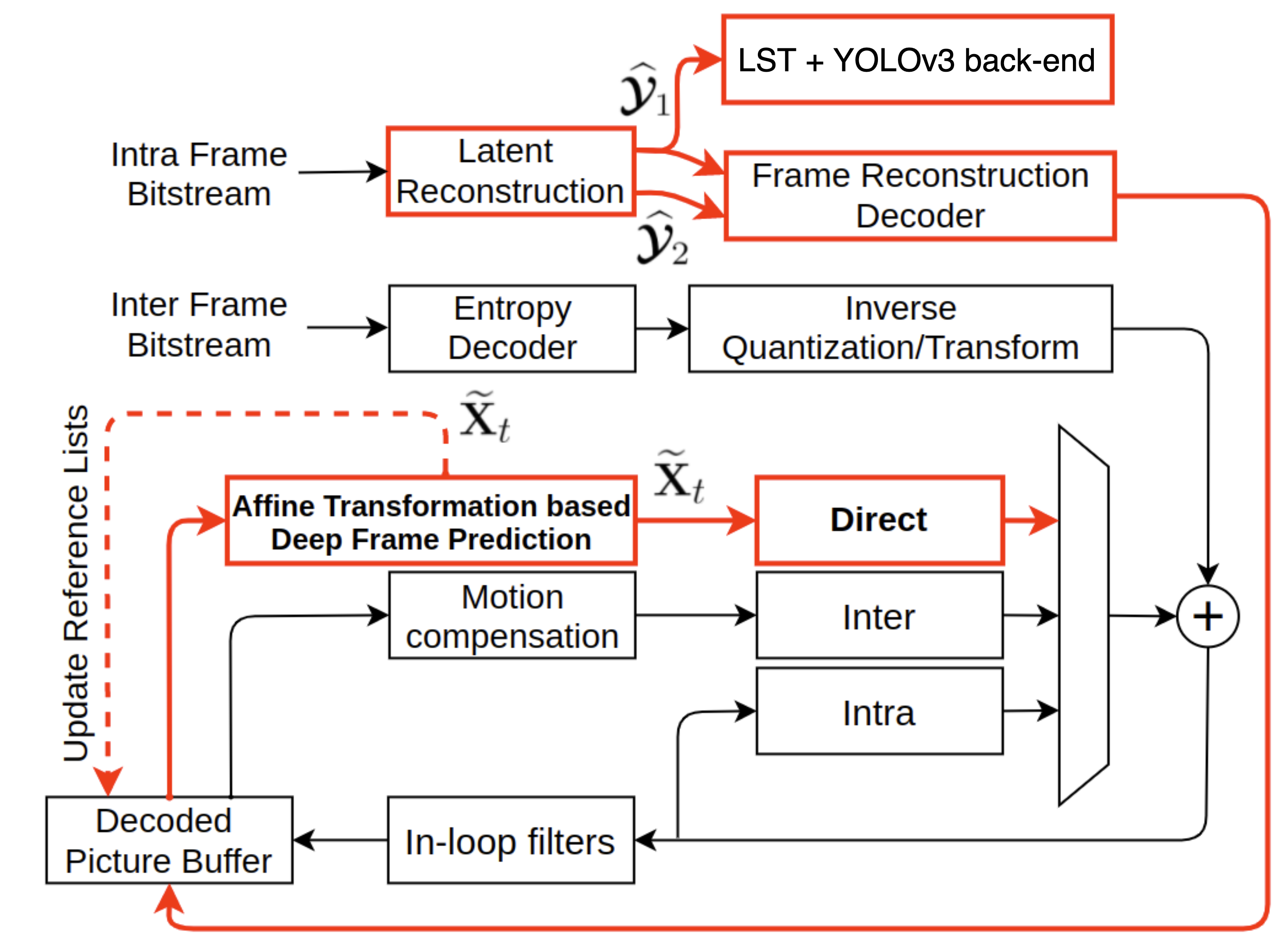}
    \end{minipage}
\caption{Block diagram of the overall video decoder. DNN-based components are shown in red.}
\vspace{-.2cm}
\label{fig:ch6_integrated_block}
\end{figure}

The structure of the overall video decoder is shown in Fig.~\ref{fig:ch6_integrated_block}. 
To identify intra frames, we still use the network abstraction layer (NAL) header with 2 bytes as in conventional HEVC. Within the header data, reserved 6-bit word to support HEVC scalability is also re-used to distinguish layer IDs for our task scalability. When only the base layer of an intra frame is received, $\widehat{\mathbfcal{Y}}_1$ is reconstructed and fed to the object detection pipeline (LST + YOLOv3 back-end) to detect objects. When the enhancement layer of an intra frame is also received,  $\widehat{\mathbfcal{Y}}_2$ can be reconstructed. Then, input frame is reconstructed from the full latent representation $\widehat{\mathbfcal{Y}}=\{\widehat{\mathbfcal{Y}}_1, \widehat{\mathbfcal{Y}}_2\}$. The reconstructed frame is registered in decoded picture buffer (DPB) so that it can be used as a reference for inter frames. 

For inter-frame coding, the coded bitstream is parsed through the entropy decoder followed by inverse quantization and transformation. Reconstructed residual signal is added to the predictor to reconstruct the input frame. For block-level prediction in the inter frame, there are conventional HEVC intra and inter prediction tools. The DNN-predicted frame can be used in the HEVC coding pipeline in several ways. One is the ``direct'' mode, in which the frame is used as an additional prediction mode, indicated by a flag bit. In this case, no additional motion information is needed. As a result, this mode ends up being selected up to 57\% of the time in the HEVC rate-distortion (RD) optimization, according to~\cite{choi2021affine}. However, this mode is constrained to use square blocks. 
For this reason, we supplement the pipeline by another approach indicated by the dashed red line in Fig.~\ref{fig:ch6_integrated_block}. Here, the frame with the largest picture order count (POC) difference from the current frame in the DPB is replaced by the DNN-generated frame $\widehat{\mathbf{X}}_t$. As such, the DNN-generated frame becomes a reference frame in the DPB. All inter-coding modes can be used in this case, but additional motion information may be required, as is the case with conventional inter-prediction. For block-level coding in inter frames, all coding modes, including the modes utilizing the DNN-generated frame $\widehat{\mathbf{X}}_t$, compete in the HEVC 
RD optimization process.

\section{Experiments}
\label{sec:ch6_experimental_results}

\subsection{Implementation and training}
All DNNs are implemented in Pytorch. Python embedding library\footnote{\url{https://docs.python.org/3/extending/}} is used to embed the DNNs into HM-16.20 implemented in C++. While compressing input video, forward operations of the DNNs to perform the multi-task compression and the frame prediction are executed across CPU and GPU.

The intra-frame codec (Fig.~\ref{fig:intra_coding}) was trained on the CLIC,\footnote{\url{http://www.compression.cc/}} JPEG AI~\cite{jpeg_ai_dataset}, and VIMEO-90K~\cite{xue2019video_vimeo} datasets, on randomly cropped patches of size $256 \times 256$, following the procedure in~\cite{choi2022scalable}. Adam optimizer with a learning rate of $10^{-4}$ was used. The codec was trained with $\gamma=0.006$ in~(\ref{eq:loss_function}), for six values of $\lambda$ shown in Table~\ref{tbl:lambda_values}. The DNN-based frame prediction network (Fig.~\ref{fig:frame_prediction}) was trained on randomly selected triplets of $152\times\ 152$ patches cropped from the videos from Xiph\footnote{\url{https://media.xiph.org/video/derf/}} and VIMEO-90K~\cite{xue2019video_vimeo}, following the procedure in~\cite{choi2021affine}. AdaMax optimizer with a learning rate of $10^{-3}$ was used.

To use the pre-trained YOLOv3~\cite{Redmon2018_yolov3} back-end in the evaluation of object-detection performance, the input resolution was resized to $512\times512$ using bilinear interpolation without letter-boxing. Compression benchmarks are the two latest video coding standards: HEVC~\cite{hevc_std} (specifically, HM-16.20) and Versatile Video Coding (VVC)~\cite{vvc_std} (specifically, VTM-10.0). We encode the test sequences using the all-intra and random access configurations with intra period of 8. For benchmarks, we encode the sequences with $\text{QP}\in\{18, 22, 26, 30, 34, 38\}$ and $\text{QP}\in\{20, 24, 28, 32, 36, 40\}$ for HEVC and VVC, respectively. For our proposed system in the all intra configuration, we encode the sequences using the six models whose $\lambda$ values are shown in Table~\ref{tbl:lambda_values}. For the random access case, to achieve the range of bit rates comparable to the benchmarks, we used several combinations of $\lambda$ values for the intra codec and QP for inter coding, as shown in Table~\ref{tbl:ch6_qp_and_lambda_for_ra}. Inter frames are coded with $\text{QP}\in\{18, 22, 26, 30, 34, 38\}$ plus the QP offsets related to the hierarchical reference structure defined in the HEVC Common Test Conditions (CTC)~\cite{hevc_ctc}. Bj{\o}ntegaard Delta (BD) metrics~\cite{bd_br,vcm_template} are used to evaluate the performance 
against the benchmarks in terms of rate-distortion and rate-accuracy.  

\begin{table}[t]
\centering
\caption{$\lambda$ values for training intra-frame coding models}
\label{tbl:lambda_values}
\begin{tabular}{@{}ccccccc@{}}
\toprule
Quality Index & 1      & 2      & 3      & 4     & 5     & 6 \\ \midrule
$\lambda$   & 0.0018 & 0.0035 & 0.0067 & 0.013 & 0.025 & 0.0483 \\ \bottomrule
\end{tabular}
\end{table}

\begin{table}[t]
\centering
\caption{Combinations of model index ($\lambda$) and QP for intra and inter frame, respectively, in random access coding }
\label{tbl:ch6_qp_and_lambda_for_ra}
\begin{tabular}{@{}c|c|cc|cc|cc@{}}
\toprule
Intra & \begin{tabular}[c]{@{}c@{}}Model index\\ ($\lambda$)\end{tabular} & \multicolumn{2}{c|}{\begin{tabular}[c]{@{}c@{}}6 \\ (0.0483)\end{tabular}} & \multicolumn{2}{c|}{\begin{tabular}[c]{@{}c@{}}5 \\ (0.025)\end{tabular}} & \multicolumn{2}{c}{\begin{tabular}[c]{@{}c@{}}4 \\ (0.013)\end{tabular}} \\ \midrule
Inter & QP                                                             & \multicolumn{1}{c|}{18}                        & 22                        & \multicolumn{1}{c|}{26}                       & 30                       & \multicolumn{1}{c|}{34}                       & 38                       \\ \bottomrule
\end{tabular}
\end{table}

\subsection{Simultaneous evaluation for human and machine vision}

First, we evaluate the performance of our system against the benchmarks simultaneously on human and machine vision. We do this on the SFU-HW-Objects-v1 dataset~\cite{choi2021dataset}, which contains COCO\footnote{\url{https://cocodataset.org}}-style object labels for a set of HEVC raw video test sequences. This dataset is also being used in MPEG-VCM~\cite{m57974}. Table~\ref{tbl:ch6_all_intra_yolov3_with_ai} summarizes the performance of our coding system versus the benchmarks, with best results indicated in bold. Since this experiment involves object detection, for which our system uses only the base layer of the intra-coded frames, the test is carried out in  the all-intra configuration. Benchmark codecs code intra frames, and decoded frames are fed to YOLOv3. In our system, only the base layer of intra frames is decoded and fed via LST to the YOLOv3 back-end, as shown in Fig.~\ref{fig:intra_coding}.

\begin{figure*}[!t]
    \centering
    \begin{minipage}[b]{0.45\linewidth}
    \centering
    \includegraphics[width=\textwidth]{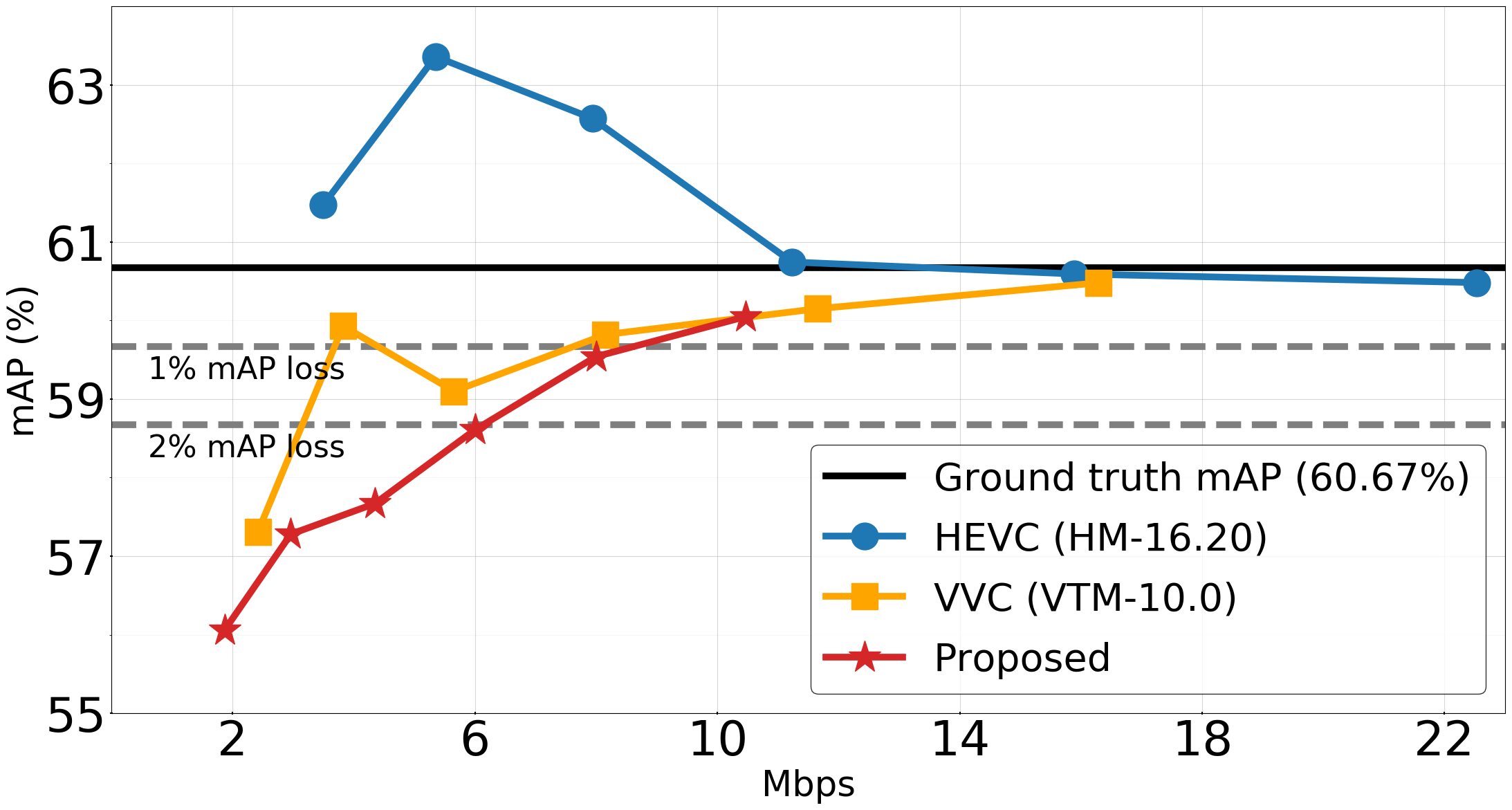}
    \centerline{(a) FourPeople}\medskip
    \end{minipage}
    \hspace{0.05\linewidth}
    \centering
    \begin{minipage}[b]{0.45\linewidth}
    \centering
    \includegraphics[width=\textwidth]{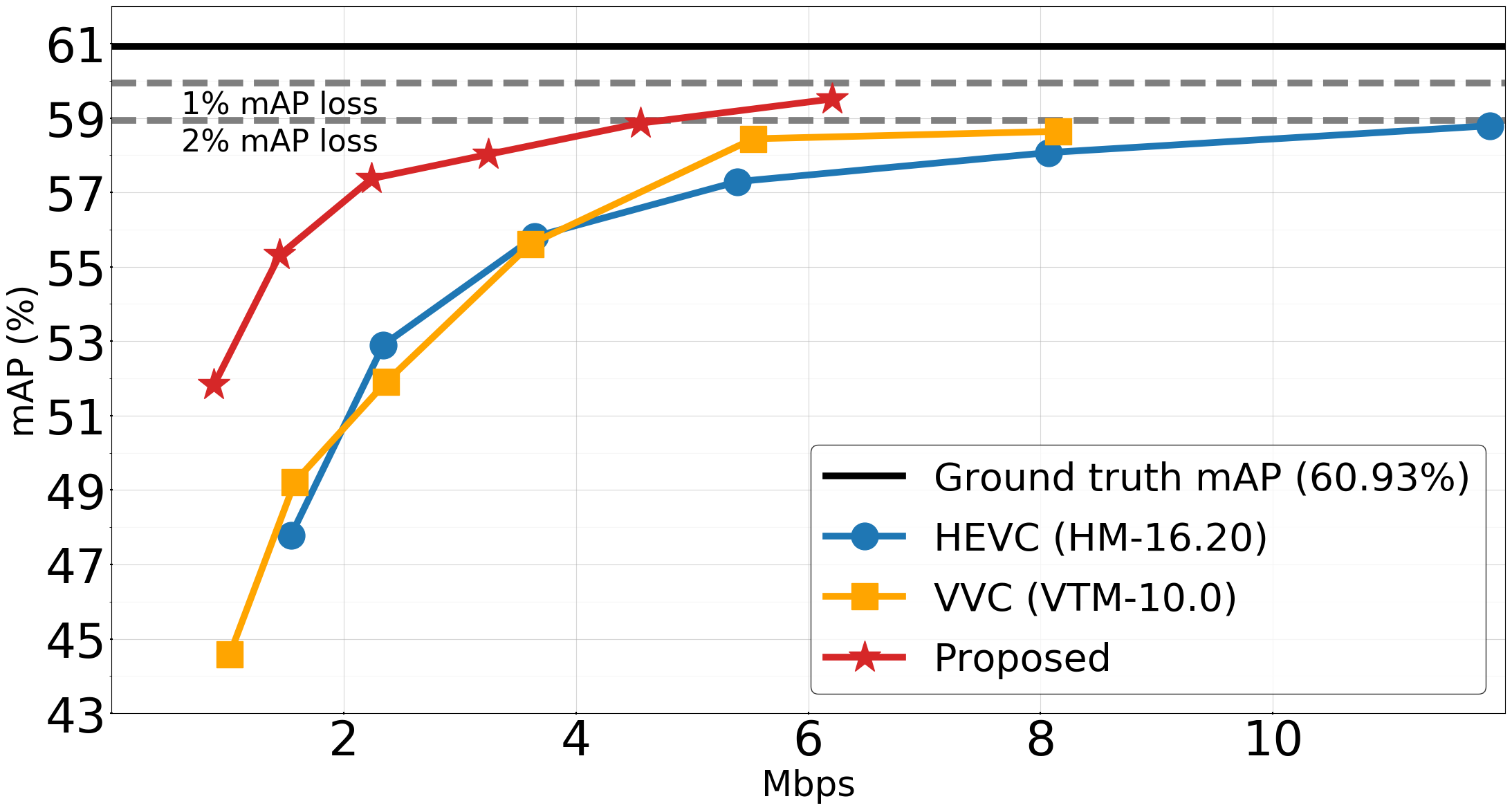}
    \centerline{(b) BasketballPass}\medskip
    \end{minipage}
\caption{Examples of rate-mAP curves: (a) shows the case when these curves are non-concave and non-convex. Moreover, there are no overlaps on the mAP axis between the blue curve and the other two curves. Hence, BD-rate-mAP cannot be reliably computed. (b) shows the case where the curves rate-mAP curves have similar characteristics as the rate-PSNR curves, so BD-rate-mAP can be reliably computed.}
\label{fig:ch6_ra_curves}
\vspace{-5pt}
\end{figure*}

Mean Average Precision (mAP)~\cite{Redmon2018_yolov3} is used as the object detection accuracy metric. Unlike the Peak Signal-to-Noise-Ratio (PSNR), mAP vs. bit rate curves are not always concave, or even monotonic~\cite{m57974}, which makes it impossible to compute a valid BD-rate-mAP value. One example is shown in Fig.~\ref{fig:ch6_ra_curves}(a) for the sequence FourPeople, where we see that HEVC and VVC curves are non-concave and non-monotonic. For this reason, the sequence FourPeople has been excluded from the results. Other sequences had well-behaved mAP vs. bit rate curves, like the one shown in Fig.~\ref{fig:ch6_ra_curves}(b) for BasketballPass.

\begin{table}[!t]
\centering
\caption{BD performance of the proposed video coding system against HEVC and VVC in the all-intra configuration}
\label{tbl:ch6_all_intra_yolov3_with_ai}
\smallskip\noindent
\resizebox{1.0\linewidth}{!}{%
\begin{tabular}{@{}c|c|ccc|ccc@{}}
\toprule
\multicolumn{2}{c|}{}                            & \multicolumn{3}{c|}{HEVC (HM-16.20)}                                                                                                                                                        & \multicolumn{3}{c}{VVC (VTM-10.0)}                                                                                                                                                           \\ \cmidrule(l){3-8} 
\multicolumn{2}{c|}{\multirow{-2}{*}{Benchmark}} & Machine Vision                                                   & \multicolumn{2}{c|}{Human Vision}                                                                                        & Machine Vision                                                    & \multicolumn{2}{c}{Human Vision}                                                                                         \\ \midrule
                        &                            & \multicolumn{3}{c|}{BD-rate-}                                                                              & \multicolumn{3}{c}{BD-rate-}                                                                              \\ \cmidrule(l){3-8} 
\multirow{-2}{*}{Class} & \multirow{-2}{*}{Sequence} & mAP                                                                    & PSNR       & MS-SSIM              & mAP                                                                    & PSNR            & MS-SSIM        \\ \midrule
\multirow{3}{*}{A} & PeopleOnStreet  & \textbf{-37.17\%}                                                     & 8.55\%                                                    & \textbf{-22.93\%}                                            & \textbf{-29.52\%}                                                      & 36.47\%                                                   & \textbf{-6.34\%}                                             \\
                   & Traffic         & 33.82\%                                                               & 16.80\%                                                   & \textbf{-20.72\%}                                            & 61.09\%                                                                & 44.38\%                                                   & \textbf{-4.09\%}                                             \\ \cmidrule(l){2-8} 
                   & Average         & \textbf{-1.68\%}                                                      & 12.67\%                                                   & \textbf{-21.83\%}                                            & 15.78\%                                                                & 40.42\%                                                   & \textbf{-5.21\%}                                             \\ \midrule
\multirow{6}{*}{B} & BQTerrace       & 16.37\%                                                               & 29.84\%                                                   & \textbf{-18.33\%}                                            & \textbf{-2.26\%}                                                       & 73.32\%                                                   & 7.84\%                                                       \\
                   & BasketballDrive & \textbf{-49.91\%}                                                     & 24.57\%                                                   & \textbf{-13.63\%}                                            & \textbf{-47.16\%}                                                      & 64.10\%                                                   & 9.47\%                                                       \\
                   & Cactus          & \textbf{-30.68\%}                                                     & 20.79\%                                                   & \textbf{-19.18\%}                                            & \textbf{-46.64\%}                                                      & 55.70\%                                                   & 2.28\%                                                       \\
                   & Kimono          & \textbf{-75.00\%}                                                     & 1.37\%                                                    & \textbf{-15.72\%}                                            & \textbf{-70.98\%}                                                      & 24.91\%                                                   & 0.74\%                                                       \\
                   & ParkScene       & \textbf{-35.81\%}                                                     & 14.63\%                                                   & \textbf{-16.45\%}                                            & \textbf{-20.30\%}                                                      & 40.05\%                                                   & \textbf{-0.63\%}                                             \\ \cmidrule(l){2-8} 
                   & Average         & \textbf{-35.01\%}                                                     & 18.24\%                                                   & \textbf{-16.66\%}                                            & \textbf{-37.47\%}                                                      & 51.62\%                                                   & 3.94\%                                                       \\ \midrule
\multirow{5}{*}{C} & BQMall          & \textbf{-51.04\%}                                                     & 1.07\%                                                    & \textbf{-20.80\%}                                            & \textbf{-51.96\%}                                                      & 31.80\%                                                   & 0.95\%                                                       \\
                   & BasketballDrill & \textbf{-37.45\%}                                                     & 0.62\%                                                    & \textbf{-22.76\%}                                            & \textbf{-46.88\%}                                                      & 46.70\%                                                   & 5.09\%                                                       \\
                   & PartyScene      & \textbf{-8.01\%}                                                      & 15.60\%                                                   & \textbf{-12.54\%}                                            & \textbf{-12.25\%}                                                      & 43.87\%                                                   & 5.33\%                                                       \\
                   & RaceHorses      & 27.07\%                                                               & 8.49\%                                                    & \textbf{-11.43\%}                                            & \textbf{-36.60\%}                                                      & 38.90\%                                                   & 8.37\%                                                       \\ \cmidrule(l){2-8} 
                   & Average         & \textbf{-17.36\%}                                                     & 6.44\%                                                    & \textbf{-16.88\%}                                            & \textbf{-36.92\%}                                                      & 40.32\%                                                   & 4.94\%                                                       \\ \midrule
\multirow{5}{*}{D} & BQSquare        & \textbf{-6.51\%}                                                      & 7.39\%                                                    & \textbf{-25.10\%}                                            & \textbf{-15.38\%}                                                      & 32.52\%                                                   & \textbf{-10.52\%}                                            \\
                   & BasketballPass  & \textbf{-57.82\%}                                                     & \textbf{-2.33\%}                                          & \textbf{-16.14\%}                                            & \textbf{-55.58\%}                                                      & 29.18\%                                                   & 6.82\%                                                       \\
                   & BlowingBubbles  & \textbf{-15.49\%}                                                     & 1.08\%                                                    & \textbf{-15.26\%}                                            & \textbf{-2.86\%}                                                       & 30.57\%                                                   & 5.72\%                                                       \\
                   & RaceHorses      & 21.69\%                                                               & \textbf{-4.15\%}                                          & \textbf{-11.10\%}                                            & \textbf{-22.45\%}                                                      & 27.46\%                                                   & 11.82\%                                                      \\ \cmidrule(l){2-8} 
                   & Average         & \textbf{-14.53\%}                                                     & 0.50\%                                                    & \textbf{-16.90\%}                                            & \textbf{-24.07\%}                                                      & 29.93\%                                                   & 3.46\%                                                       \\ \midrule
\multirow{3}{*}{E} 
                   & Johnny          & 116.35\%                                                              & 7.87\%                                                    & \textbf{-19.50\%}                                            & 86.62\%                                                                & 47.54\%                                                   & 7.45\%                                                       \\
                   & KristenAndSara  & \textbf{-39.08\%}                                                     & 7.48\%                                                    & \textbf{-29.17\%}                                            & \textbf{-8.03\%}                                                       & 42.40\%                                                   & \textbf{-8.88\%}                                             \\ \cmidrule(l){2-8} 
                   & Average         & 38.64\%           & 6.21\%                                                    & \textbf{-24.90\%}                                            & 39.29\%            & 41.19\%                                                   & \textbf{-2.60\%}                                             \\ \midrule
\multicolumn{2}{c|}{Avg. (A - D)}                         & \textbf{-20.40\%}                                                     & 9.62\%                                                    & \textbf{-17.47\%}                                            & \textbf{-26.65\%}                                                      & 41.33\%                                                   & 2.86\%                                                       \\ \midrule
\multicolumn{2}{c|}{Avg. (A - E)}                         & \textbf{-13.45\%} & 9.05\%                                                    & \textbf{-18.71\%}                                            & \textbf{-18.89\%} & 41.31\%                                                   & 1.95\%                                                       \\ \bottomrule
\end{tabular}}
\end{table}

On object detection, our coding system shows significant bit savings of 13.45\% and 18.85\% on average against HEVC and VVC, respectively, when averaged over all sequence classes.  Without Class E sequences, average bit reduction is even higher -- 20.40\% and 26.65\%, respectively, against HEVC and VVC. Surprisingly, we save more bits against VVC compared to HEVC, which implies that advanced coding tools adopted in VVC are less machine vision-friendly. 

In terms of input reconstruction for human viewing, standard codecs perform better, as expected, because that is what they are optimized for. In terms of BD-rate-PSNR, our system increases bits by about 9\% and 41\% against HEVC and VVC, respectively. In other words, the compression efficiency of VVC is far superior to other methods in terms of rate-PSNR. Meanwhile, our system performs reasonably well against HEVC, achieving some gains on two sequences in Class D. Overall, our method shows better performance in classes C and D compared to other classes. We suspect that this is due to input scaling with bilinear interpolation, which may cause some artifacts to the sequences in Class C ($832\times480)$ and D ($416\times240$) compared to sequences with higher resolution. 

In terms of BD-rate-MS-SSIM, our system outperforms HEVC by 18.71\% and has marginally worse performance (by 1.95\%) compared to VVC. If we consider MS-SSIM a more relevant metric for human viewing experience, then one could argue that our system provides comparable or better performance for human viewing while achieving gains on machine vision. Indeed, it has been known for a while that DNN-based codecs do well on MS-SSIM, and our system benefits from DNN-based intra coding in this experiment.

\subsection{Input reconstruction with random access coding}

\begin{table}[!t]
\centering
\caption{Input reconstruction performance of the proposed video coding system against HEVC  and VVC in the random access configuration with the intra period of 8}
\label{tbl:ch6_all_ra_human_vision}
\smallskip\noindent
\resizebox{1.0\linewidth}{!}{%
\begin{tabular}{@{}c|c|cc|cc@{}}
\toprule
\multicolumn{2}{c|}{Benchmark}             & \multicolumn{2}{c|}{HEVC (HM-16.20)}                                                                                                & \multicolumn{2}{c}{VVC (VTM-10.0)}  \\        \midrule
Class                 & Sequence           & \begin{tabular}[c]{@{}c@{}}BD-rate \\ (PSNR)\end{tabular} & \begin{tabular}[c]{@{}c@{}}BD-rate \\ (MS-SSIM)\end{tabular} & \begin{tabular}[c]{@{}c@{}}BD-rate \\ (PSNR)\end{tabular} & \begin{tabular}[c]{@{}c@{}}BD-rate \\ (MS-SSIM)\end{tabular}  \\ \midrule
\multirow{3}{*}{A}    & PeopleOnStreet     & \textbf{-1.27\%}                                          & \textbf{-12.15\%}                                            & 20.82\%                                                   & 9.41\%                                                        \\
                      & Traffic            & 21.88\%                                                   & 8.90\%                                                       & 48.65\%                                                   & 33.31\%                                            \\ \cmidrule(l){2-6} 
                      & Average            & 10.30\%                                                   & \textbf{-1.63\%}                                             & 34.74\%                                                   & 21.36\%                                     \\ \midrule
\multirow{6}{*}{B}    & BQTerrace          & 21.70\%                                                   & 3.32\%                                                       & 55.15\%                                                   & 32.94\%        \\
                      & BasketballDrive    & 5.85\%                                                    & \textbf{-2.02\%}                                             & 42.65\%                                                   & 31.89\%    \\
                      & Cactus             & 16.54\%                                                   & \textbf{-1.89\%}                                             & 49.58\%                                                   & 27.42\%       \\
                      & Kimono             & 0.50\%                                                    & \textbf{-9.96\%}                                             & 29.06\%                                                   & 14.88\%    \\
                      & ParkScene          & 14.13\%                                                   & 0.86\%                                                       & 39.48\%                                                   & 23.98\%       \\ \cmidrule(l){2-6} 
                      & Average            & 11.74\%                                                   & \textbf{-1.94\%}                                             & 43.18\%                                                   & 26.22\%    \\ \midrule
\multirow{5}{*}{C}    & BQMall             & 3.14\%                                                    & \textbf{-9.64\%}                                             & 40.89\%                                                   & 22.20\%   \\
                      & BasketballDrill    & 10.91\%                                                   & \textbf{-4.05\%}                                             & 56.60\%                                                   & 54.33\%         \\
                      & PartyScene         & 12.99\%                                                   & \textbf{-0.45\%}                                             & 43.24\%                                                   & 24.76\%                                  \\
                      & RaceHorses         & 4.23\%                                                    & \textbf{-1.58\%}                                             & 37.94\%                                                   & 31.42\%                             \\ \cmidrule(l){2-6} 
                      & Average            & 7.82\%                                                    & \textbf{-3.93\%}                                             & 44.67\%                                                   & 33.18\%                                 \\ \midrule
\multirow{5}{*}{D}    & BQSquare           & 7.38\%                                                    & \textbf{-9.49\%}                                             & 50.49\%                                                   & 19.02\%                                                 \\
                      & BasketballPass     & \textbf{-2.86\%}                                          & \textbf{-9.68\%}                                             & 36.77\%                                                   & 23.01\%                             \\
                      & BlowingBubbles     & 4.18\%                                                    & \textbf{-6.94\%}                                             & 39.37\%                                                   & 21.03\%                             \\
                      & RaceHorses         & \textbf{-2.71\%}                                          & \textbf{-4.75\%}                                             & 38.38\%                                                   & 31.18\%                             \\ \cmidrule(l){2-6} 
                      & Average            & 1.50\%                                                    & \textbf{-7.71\%}                                             & 41.25\%                                                   & 23.56\%                            \\ \midrule
\multirow{4}{*}{E}    & FourPeople         & 11.52\%                                                   & \textbf{-11.51\%}                                            & 45.47\%                                                   & 13.16\%                                                       \\
                      & Johnny             & 17.84\%                                                   & \textbf{-2.49\%}                                             & 62.58\%                                                   & 32.28\%                             \\
                      & KristenAndSara     & 14.26\%                                                   & \textbf{-16.50\%}                                            & 53.67\%                                                   & 11.36\%                             \\ \cmidrule(l){2-6} 
                      & Average            & 14.54\%                                                   & \textbf{-10.17\%}                                            & 53.90\%                                                   & 18.94\%                             \\ \midrule
\multicolumn{2}{c|}{Avg. (A - D)}          & 7.77\%                                                    & \textbf{-3.97\%}                                             & 41.94\%                                                   & 26.72\%                                                     \\ \midrule
\multicolumn{2}{c|}{Avg. (A - E)}          & 8.90\%                                                    & \textbf{-5.00\%}                                             & 43.93\%                                                   & 25.42\%                                             \\ \bottomrule
\end{tabular}}
\end{table}

\begin{table}[!ht]
\centering
\caption{The effect of DNN-aided frame prediction in the random access configuration}
\label{tbl:ch6_all_comp_deep_frame_prediction_or_no}
\begin{tabular}{@{}c|cc@{}}
\toprule
Class &  BD-rate-PSNR         & BD-rate-MS-SSIM              \\ \midrule
A                      & \textbf{-2.19\%}                                                   & \multicolumn{1}{c}{\textbf{-3.61\%}}                                         \\ \cmidrule(l){1-3}
B                      & 0.35\%                                                    & \multicolumn{1}{c}{0.60\%}                  \\ \cmidrule(l){1-3}
C                      & \textbf{-1.02\%}                                                   & \multicolumn{1}{c}{\textbf{-1.33\%}}                                                       \\ \cmidrule(l){1-3}
D                      & \textbf{-0.79\%}                                                   & \multicolumn{1}{c}{\textbf{-0.38\%}}                                           \\ \cmidrule(l){1-3}
E                      & \textbf{-1.33\%}                                                   & \multicolumn{1}{c}{\textbf{-1.46\%}}                                                       \\ \midrule
Average                & \textbf{-0.77\%}                                                   & \multicolumn{1}{c}{\textbf{-0.86\%}}                                                       \\ \bottomrule
\end{tabular}
\end{table}

Table~\ref{tbl:ch6_all_ra_human_vision} summarizes input reconstruction performance in terms of BD-rate metrics for the random access configuration with the intra period of 8. Here, benchmark codecs perform better than they did on the object detection task, because they were optimized for this kind of use. In terms of BD-rate-PSNR, our system increases the rate by about 8.9\% on average against HEVC. Recall that our inter-coding pipeline is built upon HEVC. Considering the fact that conventional scalable extensions of HEVC increase the bit rate by 15\%--25\% per layer~\cite{shvc_overview}, our scalable system for human and machine vision performs well within this margin. The performance against VVC in terms of BD-rate-PSNR is correspondingly lower, as expected, with about 44\% rate increase. Our codec performs much better in terms of MS-SSIM. In fact, in this case, it provides BD-rate savings of 5\%, on average, against HEVC, and the loss against VVC is now reduced to about 25\%.

\subsection{Ablation study}
Here we examine the effect of DNN-aided frame prediction within our system, by comparing the full version of the system against a stripped-down version, which does not include DNN-based frame prediction. The results are shown in Table~\ref{tbl:ch6_all_comp_deep_frame_prediction_or_no} for the random access configuration with the intra period of 8.  DNN-aided frame prediction brings 0.8\%--0.9\% bit savings on average, both in terms of PSNR and MS-SSIM.   

\subsection{Break-even points}

In earlier sections we saw that, compared with conventional HEVC or VVC coding, our system achieves compression gains when only the machine vision task is needed, but suffers coding loss in certain cases when input reconstruction is needed for human viewing. Thus, in practice, the question of whether or not our system will provide bit savings depends on how frequently input reconstruction is needed compared to machine vision. In this section, we quantify this trade-off in terms of the maximum fraction of time that input reconstruction is needed, on average, while still allowing compression gains for our system. We call this fraction of time the \emph{break-even point}.

According to Table~\ref{tbl:ch6_all_intra_yolov3_with_ai}, our system is 13.45\% more efficient (i.e., uses 0.8655 the amount of bits), on average, compared to HEVC when only object detection is required. At the same time, it is 9.05\% less efficient (i.e., uses 1.0905 the amount of bits), on average, when input reconstruction is needed, if reconstruction quality is measured by PSNR. Let $t_h \in [0,1]$ be the fraction of time that input reconstruction is needed. The amount of bits used by our system will be less than or equal to that used by HEVC if 
 \begin{equation}
(1-t_h) \cdot 0.8655 + t_h \cdot 1.0905 \le 1.
\label{eq:ch6_best_scenario_2layer_with_yolo_against_hevc}
\end{equation}
Solving for $t_h$ that achieves equality in~(\ref{eq:ch6_best_scenario_2layer_with_yolo_against_hevc}), we obtain the break-even point of $t_h =0.5978$. That is to say, if input reconstruction is needed less than 59.78\% of the time, our system will provide overall bit savings over HEVC. In fact, if input reconstruction quality is measured by MS-SSIM instead of PSNR, our system would always provide gains over HEVC, since the corresponding BD-rate-MS-SSIM is negative in Table~\ref{tbl:ch6_all_intra_yolov3_with_ai}. Repeating the same calculation for other cases, we obtain the break-even points shown in  Table~\ref{tbl:ch6_break-even_points}. The smallest break-even point across all  cases is 0.3138. Hence, if input reconstruction is needed less than 30\% of the time, our system will provide savings even against VVC.

\begin{table}[!t]
\centering
\caption{Break-even points against HEVC and VVC}
\label{tbl:ch6_break-even_points}
\smallskip\noindent
\resizebox{0.75\linewidth}{!}{%
\begin{tabular}{@{}c|cc|cc@{}}
\toprule
Benchmark                                             & \multicolumn{2}{c|}{HEVC} & \multicolumn{2}{c}{VVC} \\ \midrule
Metric & PSNR        & MS-SSIM     & PSNR       & MS-SSIM    \\ \midrule
Break-even point                                                               & 0.5978      & 1.0         & 0.3138     & 0.9064     \\  \bottomrule    
\end{tabular}}
\end{table}

\section{Conclusion}
\label{sec:ch6_conclusion}

We developed a new scalable video coding system supporting machine vision (object detection) in the base layer and input reconstruction for human viewing in the enhancement layer. The system was benchmarked against the two most recent video coding standards - HEVC and VVC. The results show that the proposed system provides savings of  13-19\% on the object detection task. At the same time, it provides comparable or better performance in terms of MS-SSIM in the all-intra coding configuration, as well as the random access configuration against HEVC. Against VVC in the random access configuration, there is a loss of about 44\% in terms of BD-rate-PSNR and about 25\% in terms of BD-rate-MS-SSIM. But even against VVC, our system provides rate savings, so long as human viewing is needed less than 30\% of the time.

\bibliographystyle{IEEEtran}
\bibliography{refs}

\end{document}